\documentclass{IEEEtaes}

\usepackage{color,array,amsthm}
\usepackage{graphicx}

\jvol{XX}
\jnum{XX}
\jmonth{XXXXX}
\paper{1234567}
\pubyear{2024}
\doiinfo{TAES.2024.Doi Number}

\usepackage{amssymb,amsfonts}
\usepackage{bm}
\usepackage{algorithmic}
\usepackage{graphicx}
\usepackage{textcomp}
\usepackage{color}  

\usepackage{amsmath}
 \renewcommand{\boldsymbol}{\mathbf}

\newcommand{\cp}[1]{\ifmmode {\mathcal{#1}}\else ${\mathcal{#1}}$\fi}
\usepackage{array}
\newcolumntype{P}[1]{>{\centering\arraybackslash}p{#1}}

\newcommand{\bB}{\boldsymbol{B}}

\newcommand{\bD}{\boldsymbol{D}}

\newcommand{\bF}{\boldsymbol{F}}

\newcommand{\bH}{\boldsymbol{H}}
\newcommand{\bI}{\boldsymbol{I}}

\newcommand{\bP}{\boldsymbol{P}}
\newcommand{\bQ}{\boldsymbol{Q}}
\newcommand{\bR}{\boldsymbol{R}}

\newcommand{\bU}{\boldsymbol{U}}

\newcommand{\bb}{\boldsymbol{b}}

\newcommand{\bbm}{\boldsymbol{m}}

\newcommand{\bp}{\boldsymbol{p}}

\newcommand{\br}{\boldsymbol{r}}

\newcommand{\bs}{\boldsymbol{s}}

\newcommand{\bv}{\boldsymbol{v}}
\newcommand{\bx}{\boldsymbol{x}}

\newcommand{\bz}{\boldsymbol{z}}

\newcommand{\bSigma}{\boldsymbol{\Sigma}}

\usepackage{nccmath}

\definecolor{darkgreen}{rgb}{0., 0.4, 0.}
\definecolor{amber}{rgb}{1.0, 0.49, 0.0}
\definecolor{orange}{rgb}{1.0, 0.4, 0.0}

\usepackage{xcolor}
\usepackage{amsmath,amsfonts}
\newcommand{\norm}[1]{\left\lVert#1\right\rVert}
\usepackage{subfig}
\usepackage{algorithm2e}
\usepackage{mathtools}
\begin{document}
\title{Mitigation of Radar Range Deception Jamming Using Random Finite Sets} 

%
\author{Helena Calatrava}
\member{Student Member, IEEE}
\affil{Northeastern University, Boston, MA 02115, USA} 
\author{Aanjhan Ranganathan}
\member{Member, IEEE}
\affil{Northeastern University, Boston, MA 02115, USA} 
\author{Tales Imbiriba}
\member{Member, IEEE}
\affil{Northeastern University, Boston, MA 02115, USA} 
\author{Gunar Schirner}
\member{Member, IEEE}
\affil{Northeastern University, Boston, MA 02115, USA} 
\author{Murat Akcakaya}
\member{Senior, IEEE}
\affil{Northeastern University, Boston, MA 02115, USA} 
\author{Pau Closas}
\member{Senior, IEEE}
\affil{Northeastern University, Boston, MA 02115, USA} 
%
%
\receiveddate{Manuscript received XXXXX 00, 0000; revised XXXXX 00, 0000; accepted XXXXX 00, 0000. This work was supported by the Army Research Laboratory under Cooperative Agreement Number W911NF-23-2-0014, and in part by the National Science Foundation under Awards ECCS-1845833 and CCF-2326559.}
\corresp{{\itshape (Corresponding author: Helena Calatrava)}.}

\authoraddress{Helena Calatrava, Aanjhan Ranganathan, Tales Imbiriba, Gunar Schirner, and Pau Closas are with Northeastern University, Boston, MA 02115, USA
(e-mail: \{calatrava.h, aanjhan, t.imbiriba, g.schirner, closas\}@northeastern.edu). Murat Akcakaya is with University of Pittsburgh, PA 15260, USA (e-mail: akcakaya@pitt.edu)}
%
%
%
\markboth{CALATRAVA ET AL.}{MITIGATION OF RADAR RANGE DECEPTION JAMMING USING RANDOM FINITE SETS}
\maketitle
%
%
\begin{abstract}This paper presents a radar target tracking framework for addressing main-beam range deception jamming attacks using random finite sets (RFSs). Our system handles false alarms and detections with false range information through multiple hypothesis tracking (MHT) to resolve data association uncertainties. We focus on range gate pull-off (RGPO) attacks, where the attacker adds positive delays to the radar pulse, thereby mimicking the target trajectory while appearing at a larger distance from the radar. The proposed framework incorporates knowledge about the spatial behavior of the attack into the assumed RFS clutter model and uses only position information without relying on additional signal features. We present an adaptive solution that estimates the jammer-induced biases to improve tracking accuracy as well as a simpler non-adaptive version that performs well when accurate priors on the jamming range are available. Furthermore, an expression for RGPO attack detection is derived, where the adaptive solution offers superior performance. The presented strategies provide tracking resilience against multiple RGPO attacks in terms of position estimation accuracy and jamming detection without degrading tracking performance in the absence of jamming.
\end{abstract}

\section{INTRODUCTION}

%
%

\begin{figure*}
\centering
  \includegraphics[width=0.95\textwidth]{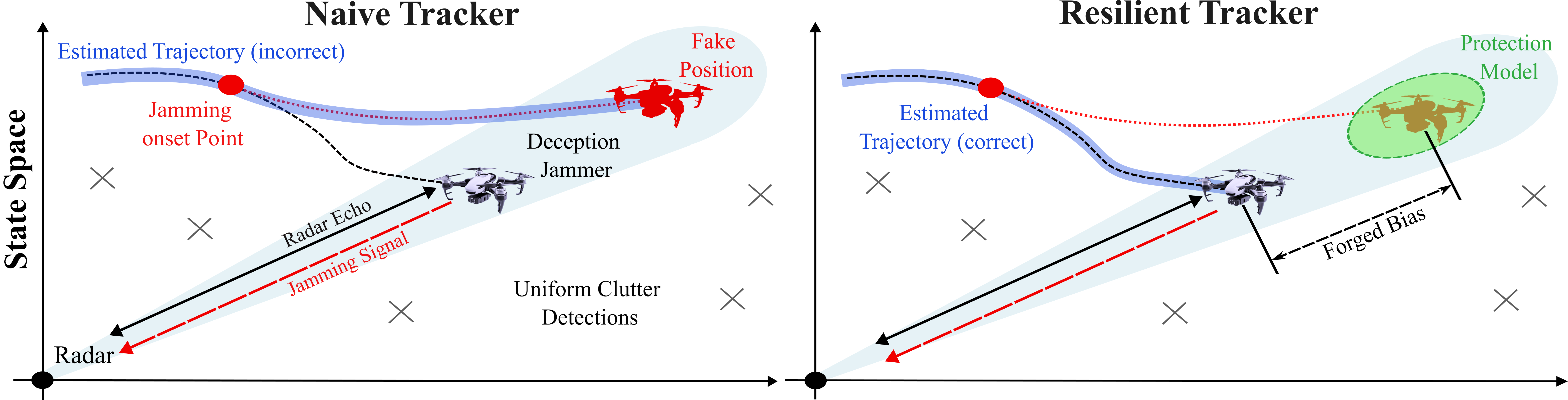}
  \vspace{-.05cm}
  \caption{Comparison of radar tracking without safeguards (naive tracker) and with protection mechanisms (resilient tracker) against main-beam deception jamming. The naive tracker is misled by jamming, while the resilient tracker estimates and corrects the jammer-induced bias in the target range, thus ensuring accurate target tracking.}
   \label{fig:intro}
\end{figure*}

%
%
%
Deception jammers, also known as repeater jammers, are typically used as a self-protection strategy by systems such as tactical aircraft operating in environments with a high density of enemy radar systems.
These jammers disrupt the focus of target tracking radars (TTR) by intercepting, modifying, and retransmitting the signal of interest with false information, thereby diverting attention away from the actual target of interest (TOI)~\cite{fundamentals_ew, ew_101, 395232}.
%
%
Main-beam range deception jamming may also occur when the jammer is co-located with the TOI, or at the same angle relative to the TTR, thus acting as a false target generator~\cite{wang_main-beam_2020}.
While more power-efficient than noise jammers, deception jammers rely on memory as their critical component. In particular, digital RF memory (DRFM) technology is used to monitor and store radar signals for accurate replay attacks~\cite{roome1990digital}.
Figure~\ref{fig:intro} contrasts target tracking with (resilient tracker) and without (naive tracker) protective measures. The resilient tracker corrects the bias introduced by the attacker in the target range.
%

Range gate pull-off (RGPO) is a self-protection strategy where the attacker uses cover pulses to capture the range gate used for TOI selection and adds delays to shift it away from the target. Once it has moved significantly, the jammer shuts down and the TTR is forced to restart its search~\cite{fundamentals_ew,blair_benchmark_1998}.
Efforts in the literature focus on optimizing RGPO strategies for track deception.
The intricate nature of jammer-radar interactions complicates quantitative optimization and leads to the exploration of black-box RGPO jamming, where the jammer lacks knowledge of the TTR tracking model~\cite{10032699, jia_intelligent_2020}, unlike in the white-box scenario~\cite{9776498}.
Range gate pull-in (RGPI) attacks are typically considered impractical~\cite{6159736} due to the assumption that the TTR employs waveform diversity and pulse agility strategies~\cite{zhang_new_2013, akhtar_eccm_2007}, including the use of random OFDM signals~\cite{schuerger2009performance}. Given this, we focus on RGPO attacks. 

Conventional anti-jamming strategies use interference feature discrimination or ensure track continuity by considering the TOI motion state~\cite{he_feature-aided_2023}.
One possible approach is to leverage that deceptive measurements often have nearly identical angles to true target measurements~\cite{lan2020suppression}, allowing for the identification of deception based on small angular differences between measurement pairs~\cite{1995_slocumb,1999_li}.
%
%
%
The study in~\cite{4472184} takes advantage of a spatial feature where the steering vector of the deception jammer aligns on a cone centered around the TOI steering vector.
Furthermore, the amplitude difference between cover and target return pulses has proven informative, potentially enhancing tracking accuracy~\cite{7377013}.
Nonetheless, these methods based on low-order statistics may be insufficient when jamming signals and true targets share similar features.
%
%
%

%

%
Multiple hypothesis tracking (MHT) is the leading method for addressing the data association (DA) problem in modern target tracking systems~\cite{1102177,1263228} since it considers multiple hypotheses about the TOI state and updates them as new measurements are received.
Although MHT provides track continuity, RGPO identification still requires a decision-making process. This can be guided by heuristic methods, such as assuming that larger ranges correspond to false targets. For instance, in~\cite{1998_Kirubarajan} they reduce the association probabilities of measurements at farther ranges. However, these assumptions can lead to significant errors or track loss, especially in the context of RGPI attacks or false alarm measurements~\cite{766953}. Building on the previous paragraph, some MHT-based approaches also use signal features, such as amplitude information, to improve deception identification~\cite{hou_multiple_nodate}.

To the best of the authors' knowledge, the body of literature addressing range deception jamming through target tracking algorithms like MHT is limited, with most existing studies either relying on feature extraction, requiring multiple radar systems, or depending on the previously mentioned heuristic assumptions.
In contrast, our work relies on motion state information and incorporates knowledge of the spatial behavior of RGPO attacks into the clutter model assumed by the tracker.
This is made possible through the use of random finite sets (RFSs)~\cite{mahler_statistical_2007}, which offer a mathematically elegant tool for modeling measurement sets with variable cardinality~\cite{9353973} and are useful in the presence of detection uncertainty and false alarms.
%
%
%
%
The main contributions of the paper are as follows:
\vspace{-0.15cm}
\begin{itemize}
    \item A feature-independent target tracking solution resilient to range deception jamming that incorporates the spatial characteristics of RGPO attacks into the RFS clutter model. We present an adaptive approach for estimating jammer-induced biases and a non-adaptive approach for scenarios where accurate priors on deceptive ranges are available.  
    \item A method to monitor the likelihood of deception jamming exposure for the TTR at each time step, enabling RGPO attack detection.
    \item A novel approach to dynamically manage mixture components from the RFS clutter model that enables the mitigation of simultaneous RGPO attacks.
\end{itemize}
We evaluate the proposed strategies across four scenarios in two experiments: one involving up to one RGPO attack at a time and the other involving simultaneous attacks, with each experiment covering both straight-line and maneuvering target trajectories. Performance is compared to a clairvoyant tracker operating without DA uncertainty, and a naive tracker that only addresses false alarms. Additionally, a loss of efficiency (LoE) experiment assesses tracking performance without jamming using the posterior Cramér-Rao bound (PCRB)~\cite{668800} as a benchmark.

The remainder of the paper is structured as follows: Section II describes the system model, including the assumptions and framework of the tracking algorithm; Section III details the proposed MHT-based jamming mitigation and detection approach; Section IV presents the experimental setup and simulation results; and Section V concludes our work.
\vspace{-0.39cm}

\section{SYSTEM MODEL AND ATTACK VECTOR}
In this section, we describe the general framework of our study, the assumptions made by the TTR regarding target motion and attack vectors, and the model used for attack generation.
\vspace{-0.5cm}
\subsection{General Framework}
We consider a monostatic radar system tasked with tracking a TOI that is either co-located with a deception jammer or functions as a jammer itself. The jammer is equipped with DRFM capabilities.
In our scenario, detections can originate from the target, the deception jammer, and false alarms (uniform clutter). The latter includes reflections from buildings, trees, the ground, weather phenomena, and other objects in the environment.
%
%
For each object, at most one measurement per time step is assumed.
The possibly multiple false target detections may be referred to as deceptive measurements or secondary target-generated measurements, as they depend on the TOI state. This is similar to spawned targets, except that spawned trajectories evolve independently of their parent target~\cite{garcia-fernandez_tracking_2022,bryant_generalized_2018}.
In the proposed scenario, the target is assumed to be continuously present, whereas the attack may appear or disappear. Both primary and secondary target-generated measurements are independently intercepted with a certain detection probability.

Raw readings in radar systems typically consist of range and Doppler measurements, which are inherently non-linear, rather than direct 2D positional data. These measurements are usually converted into Cartesian coordinates by combining range data with the known orientation and position of the radar. This conversion is performed without loss of generality and results in a linear observation model, which eases the integration of multiple sensor readings and simplifies the visualization of target paths.
Considering this, the TOI state vector includes the 2D position $\bp_k = [x_k,\ y_k]^\top$ and velocity $\bv_k = [\dot{x}_k,\ \dot{y}_k]^\top$ as $\bx_k = [\bp_k^\top, \bv_k^\top]^\top$. We assume a radar located at the origin of coordinates $\bp_k^0 = [x_k^0,\ y_k^0]^\top$, which gives the line of sight (LOS) vector $\br_k = \frac{\bp_k - \bp_k^0}{\norm{\bp_k - \bp_k^0}}$ at time step $k$.
\vspace{-0.38cm}
\subsection{Attack Description}
%
%

%
Since the target and the jammer are assumed to be either co-located or the same entity, the echoes sent by the jammer are along the LOS direction. Consequently, the TTR assumes they originate from the true target. This is depicted in Figure~\ref{fig:intro}. 
We focus on linear RGPO attacks, where the deceptive measurement range is given by~\cite{blair_benchmark_1998}
\begin{equation}\label{eq:rgpo}
    R_k^{d}=R_k^t + v_\text{po}(t_k - t_0),
\end{equation}
being $R_k^{d}$ and $R_k^t$ the ranges of the deceptive and true target measurements at time step $k$, $v_\text{po}$ the attack pull-off velocity, $t_k$ the radar dwell time, and $t_0$ the attack starting time.
It is assumed that the times of arrival of the real target and jamming returns differ more than the radar resolution and consequently the two returns can be resolved. 
%

\begin{figure}[h]
    \centering
    \subfloat[Scenario 1: Straight Line + RGPO]{
        \includegraphics[height=2.35cm,width=.455\textwidth]{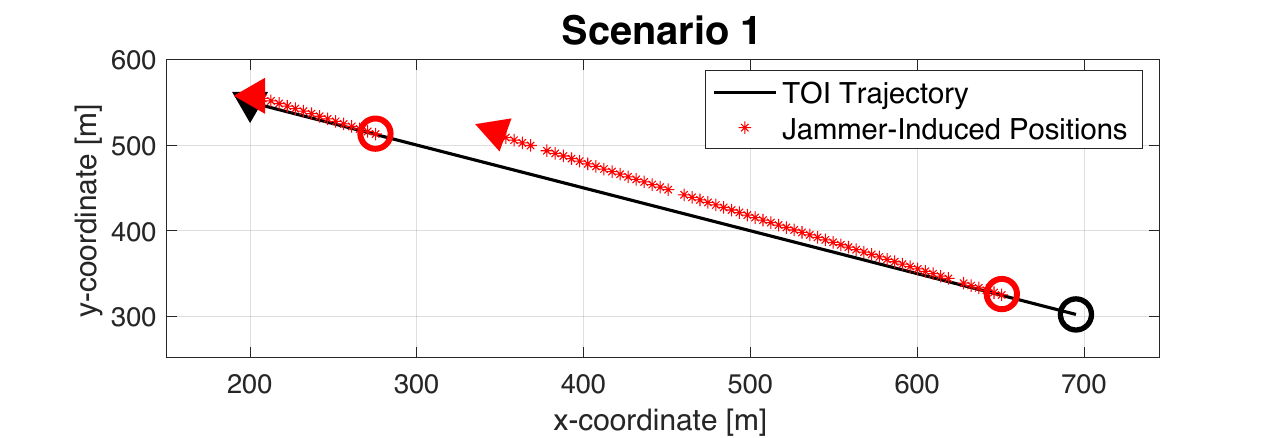}
        \label{fig:sub1}
    }
    \vspace{-0.2cm}

    \subfloat[Scenario 2: Turns + RGPO]{
        \includegraphics[height=2.35cm,width=.455\textwidth]{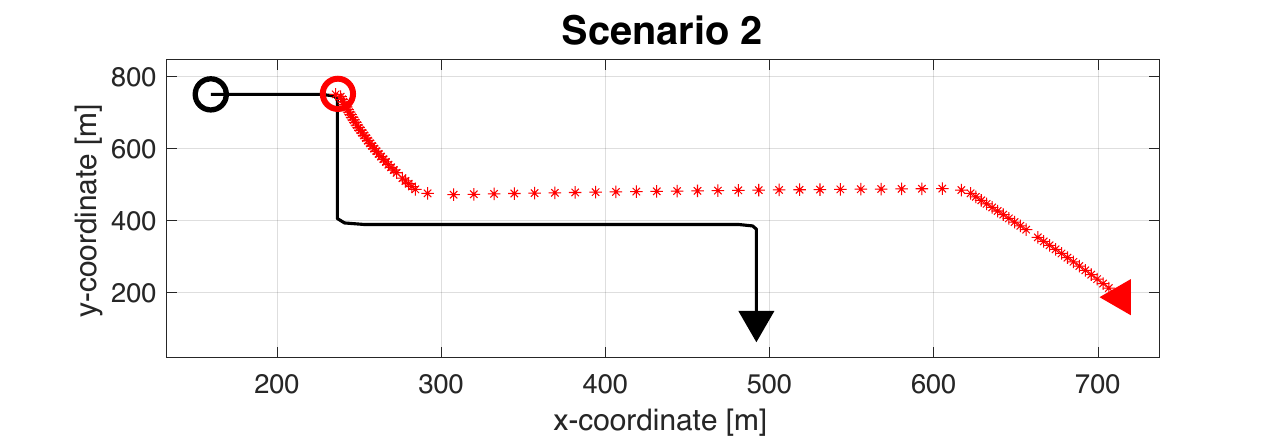}
        \label{fig:sub2}
    }
    \vspace{-0.2cm}

        \subfloat[Scenario 3: Straight Line + Multiple RGPOs]{
        \includegraphics[height=2.35cm,width=.455\textwidth]{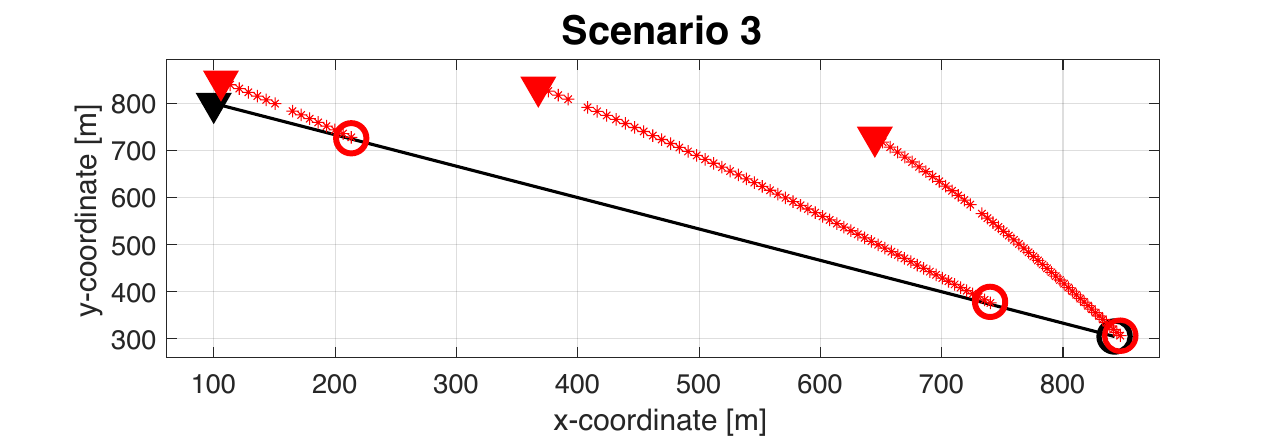}
        \label{fig:sub2}
    }
    \vspace{-0.2cm}

    \subfloat[Scenario 4: Turns + Multiple RGPOs]{
        \includegraphics[height=2.35cm,width=.455\textwidth]{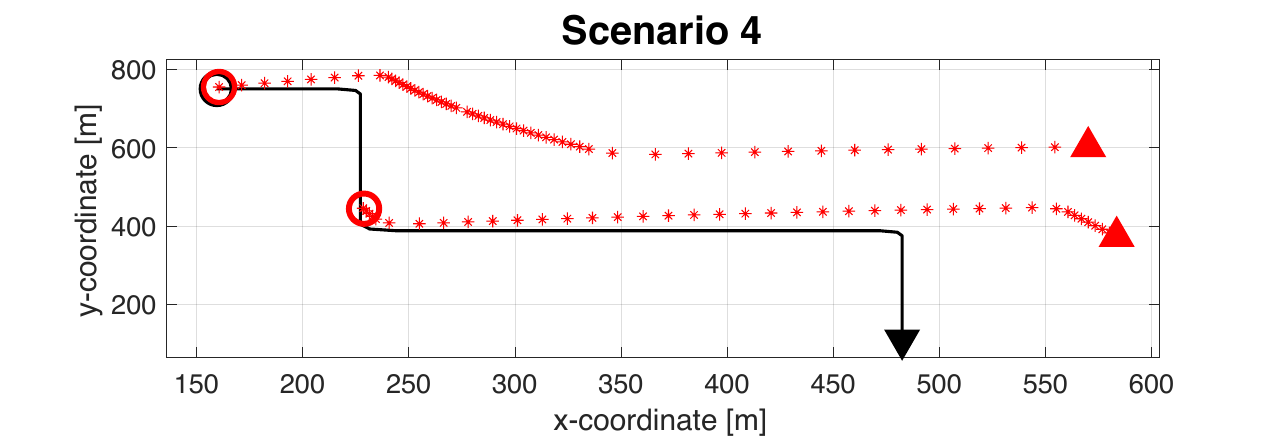}
        \label{fig:sub3}}
    \caption{TOI trajectory and jammer-induced positions for the four scenarios under study. The start and end of each trajectory are marked by circle and triangle markers, respectively. The radar is located at the origin of coordinates.}
    \label{fig:trajectories}
\end{figure}

%
To pose a threat to the radar with random jumps in position, the jammer must transmit a large number of pulses in the same pulse repetition interval, as proposed in~\cite{wan_range_2022}. This requires sophisticated equipment and precise timing and is ineffective in replicating the consistent motion pattern of a real target.
The first experiment focuses on single-return jamming attacks. This type of attack requires fewer resources to generate than multi-pulse strategies and is therefore more common.
The proposed method for handling single-return attacks also effectively manages multiple returns when they are in close proximity. Additionally, in the second experiment, we extend the method to enhance resilience against simultaneous RGPO attacks, even when their trajectories differ significantly due to variations in starting times or pull-off velocities.
The TOI trajectories and the jammer-induced positions in the four scenarios under study, spanning $100$ time steps, are depicted in Figure~\ref{fig:trajectories}. %
In Experiment 1, the target follows a straight-line trajectory (Scenario 1) or makes three 3 g turns (Scenario 2) in the presence of at most one linear RGPO attack at a time. Specifically, in Scenario 1, the first RGPO attack starts at \( k = 10 \) and continues until \( k = 75 \); then a second attack starts at \( k = 85 \) and continues indefinitely.
In Scenario 2, only one RGPO attack occurs at the start of a 3 g turn, beginning at \( k = 10 \) and continuing indefinitely.
In Experiment 2, the target follows a straight-line trajectory (Scenario 3) or makes three 3 g turns (Scenario 4) in the presence of multiple linear RGPO attacks. In Scenario 3, two simultaneous RGPO attacks occur: the first starting at \( k = 1 \) and continuing until \( k = 50 \) and the second starting at \( k = 15 \) until \( k = 75 \). A third RGPO attack starts at \( k = 85 \) and continues indefinitely.
Finally, Scenario 4 starts with a linear RGPO until \( k = 60 \). A last attack occurs at the start of a 3 g turn, beginning at \( k = 40 \) and continuing until \( k = 80 \).
%

%
%
\vspace{-1cm}

\subsection{Signal Model}\label{sec:signal_model}
In classical Bayesian filtering, the hidden state $\bx_k$ follows a first-order Markov process on the state space $\mathcal{X}\subseteq\mathbb{R}^{d_x}$ described by the transition density $f_{k|k-1}(\bx_k|\bx_{k-1})$. The radar partially observes this process in the space $\mathcal{Z}\subseteq\mathbb{R}^{d_z}$ modeled by the measurement likelihood function $g_k(\bz_k|\bx_k)$. The observation $\bz_k$ is conditionally independent of the measurement and state histories given the state $\bx_k$.
We assume a linear Gaussian transition and measurement likelihood as
\begin{equation}\label{eq:transition_model}
    f_{k|k-1}(\bx_k|\bx_{k-1}) = \mathcal{N}(\bx_k; \bF_{k-1}\bx_{k-1}, \bQ_{k-1})
\end{equation}
\begin{equation}\label{eq:measurement_likelihood}
    g_k(\bz_k|\bx_k) = \mathcal{N}(\bz_k;\bH_k\bx_k, \bR_k),
\end{equation}
where $\bF_{k-1}$ is the transition matrix of the target dynamic model and $\bH_k$ is the measurement model matrix. The covariances of the transition and measurement models are represented by $\bQ_{k-1}$ and $\bR_k$, respectively.
\subsubsection{RFS Measurement Model}
Traditional filtering methods operate assuming exactly one target-generated measurement and the absence of clutter, with measurements defined as random vectors on $\mathcal{Z}$. Conversely, the RFS framework accounts for multiple target-generated measurements, detection uncertainty and false alarms.
At time $k$, the radar receives an unordered set of $n_k = |Z_k|$ measurements $Z_k = \{\bz_{k,1}, \hdots, \bz_{k,n_k}\}$ defined on the space of finite subsets of $\mathcal{Z}$, denoted as $\mathcal{F}(\mathcal{Z})$. The RFS measurement equation is given by 
\begin{equation}
    Z_k = \Theta_k(\bx_k) \cup J_k(\bx_k) \cup W_k,
\end{equation}
where $\Theta_k(\bx_k)$ is the RFS of the primary target-generated measurement, $J_k(\bx_k)$ is the RFS of the (possibly multiple) secondary target-generated measurements, and $W_k$ is the state-independent RFS accounting for false alarm detections. $\Theta_k(\bx_k)$ is modeled as a binary RFSs
\begin{equation}
    \Theta_k(\bx_k) = \begin{cases}
             \emptyset  & \text{with probability } 1-p_D\\
             \{\bz_k\}  & \text{with prob. density } g_k(\bz_k|\bx_k)p_D,
       \end{cases}
\end{equation}
where $p_D$ is the probability of detection for the primary target-generated measurement, considered to be time-invariant and state-independent due to the constant sensor field of view (FOV) assumption.
\subsubsection{TTR Clutter Model Assumption}\label{sec:model_ttr}
%
%
The two sets of secondary target-generated measurements are grouped as the union of statistically independent Poisson RFSs as
\begin{equation}\label{eq:clutter_model}
    K_k(\bx_k) = J_k(\bx_k) \cup W_k,
\end{equation}
with intensity function $\lambda_{K,k}(\cdot|\bx_k) = \lambda_{J,k}(\cdot|\bx_k) + \lambda_{W,k}(\cdot)$. The Poisson RFS model typically assumes a variable number of detections that are generated independently. The false alarm detections are assumed to be uniformly distributed over the sensor FOV as
\begin{equation}\label{eq:likelihood_clutter}
   \lambda_{W,k}(\bz_k) = \bar{\lambda}_{0,k}u(\bz_k),
\end{equation}
where $u(\bz)$ is the uniform probability density over $\mathcal{Z}$ and $\bar{\lambda}_{0,k}$ is the expected number of uniform clutter detections.
%
The tracker assumes a linear Gaussian intensity of the secondary target-generated measurements as 
\begin{equation}\label{eq:intensity_jammer}
   \lambda_{J,k}(\bz_k|\bx_k) = \bar{\lambda}_{1,k}c_{1,k}(\bz_k|\bx_k),
\end{equation}
\begin{equation}
\label{eq:likelihood_ck1}
   c_{1,k}(\bz_k|\bx_k) = \mathcal{N}(\bz_k; \bB_k\bx_k+\bb_k,\bD_k),
\end{equation}
where $\bar{\lambda}_{1,k}$ is the expected number of jammer-induced returns. The state is observed through matrix \( \bB_k \) with a bias \( \bb_k \) along the LOS direction and observation noise covariance \( \bD_k \).
%
%
The probability of $K_k(\bx_k)$ having $n_k$ measurements is $\rho_{K,k}(n_k|\bx_k) = (\rho_{W,k}\ast\rho_{J,k})(n_k|\bx_k)$, where $\ast$ denotes convolution. This cardinality distribution is Poisson with rate $\bar{\lambda}_{0,k} + \bar{\lambda}_{1,k}$, being $\rho_{W,k}(\cdot)$ and $\rho_{J,k}(\cdot|\bx_k)$ the cardinality distributions of $W_k$ and $J_k(\bz_k|\bx_k)$. The individual elements of $K_k(\bx_k)$ are independent and identically distributed (IID) following the probability density~\cite{vo_bayesian_2008}
\begin{equation}\label{eq:clutterlikelihood}
   c_k(\bz_k|\bx_k) = w_{0,k}u(\bz_k) + w_{1,k}c_{1,k}(\bz_k|\bx_k),
\end{equation}
where $w_{i,k} = \bar{\lambda}_{i,k}/(\bar{\lambda}_{0,k}+\bar{\lambda}_{1,k})$ is the normalized weight for the density of the $i$-th secondary set of measurements.
\subsection{Interference Generation}
%
%

We consider RGPO attacks that may occur simultaneously, but each produces at most one return. Rather than a Poisson RFS this attack is modeled as a binary RFS:
\begin{equation}
    J_k^\ast(\bx_k) = \begin{cases}
             \emptyset  & \text{with probability } 1-p_J\\
             \{\bz_k\}  & \text{with prob. density } c^\ast_{1,k}(\bz_k|\bx_k)p_J,
       \end{cases}
\end{equation}
where $p_J$ is the probability of detecting a deceptive return, which is time-invariant and state-independent given the constant FOV assumption, analogous to $p_D$. Here, $c^\ast_{1,k}(\cdot|\bx_k)$ denotes the density used to generate the jammer-induced measurements.

Considering that deceptive measurements are generated as a binary RFS, the TTR assumption that the combined jammer and false alarm measurements follow a Poisson RFS leads to a model mismatch, denoted as $\ast$. Nonetheless, successful interference mitigation is still achieved by integrating the RGPO spatial characteristics into the likelihood model in~\eqref{eq:likelihood_ck1}, as explained in the next section. Furthermore, the dynamic estimation of the jammer-induced bias in the proposed adaptive strategy enhances TTR robustness, particularly in scenarios where the ranges of attack cannot be anticipated, which is often the case.
Additionally, the Poisson RFS assumption allows the single-return countermeasure to mitigate multiple attacks when their starting times and pull-off velocities are similar, even before extending the methodology to handle multiple attacks. 

Since the interference is not generated using a Poisson RFS, the parameter \(\bar{\lambda}_{1,k}\) no longer represents the average number of jamming returns per scan.
Nevertheless, we provide an interesting interpretation of this parameter by viewing the normalized weights in~\eqref{eq:clutterlikelihood} as an indicator of whether a measurement in \( K_k(\bx_k) \) is clutter-generated (\(i=0\)) or jammer-generated (\(i=1\)). Higher values of \(\bar{\lambda}_{1,k}\) are associated with a higher likelihood that the tracker will classify measurements in \( K_k(\bx_k) \) as jammer-generated.

%

%
%
%
%
\section{MHT-BASED DECEPTION JAMMING MITIGATION AND DETECTION}
%
This section introduces the proposed strategies for mitigating radar range deception jamming using MHT to resolve the DA problem.
%
%
We present a method for managing mixture components in the RFS clutter model to mitigate multiple RGPO attacks and a technique for deception jamming detection. Implementation notes on model computational complexity are also included.
%
\subsection{Background}
This section reviews the use of RFSs for Bayesian optimal single-target tracking with set-valued measurements under the linear Gaussian assumption.
According to~\cite[Proposition 1]{vo_bayesian_2008} and considering the previously stated system assumptions, the probability density for the set-valued measurement $Z_k$ is given by
\begin{equation}\label{eq:likelihood_rfs}
\begin{split}
    \eta_k(Z_k|\bx_k) &\propto (1-p_D)\cdot\rho_{K,k}(|Z_k|)\cdot|Z_k|! \\ &\times\prod_{\bz_k\in Z_k}c_k(\bz_k|\bx_k) + p_D \\
    &\times\rho_{K,k}(|Z_k - 1|)\cdot(|Z_k|-1)! \\ &\times\sum_{\bz'_k\in Z_k}g_k(\bz'_k|\bx_k)\prod_{\bz_k\neq \bz'_k}c_k(\bz_k|\bx_k),
\end{split}
\end{equation}
where the first summand corresponds to the target misdetection hypothesis where $Z_k=K_k(\bx_k)$, and the remaining $|Z_k|$ summands ($|\cdot|$ denotes the cardinality of a set) correspond to the target detection hypotheses where $\Theta_k(\bx_k) = \{\bz'_k\}$ is the primary target-generated measurement and $K_k(\bx_k)=Z_k\backslash\{\bz_k'\}$. The factorial terms account for the permutation of measurements.  
%
%

%
The predicted density is the marginal distribution of the state $\bx_k$ given the set-valued measurements up to time $k-1$ (denoted as $Z_{1:k-1} = \{Z_1,\hdots,Z_{k}\}$). This distribution is obtained by the Chapman-Kolmogorov equation as
\begin{equation}\label{eq:chapmankolmogorov_rfs}
\begin{split}
    &p_{k|k-1}(\bx_k|Z_{1:k-1}) \\ &= \int f_{k|k-1}(\bx_k|\bx_{k-1})p_{k-1|k-1}(\bx_{k-1}|Z_{1:k-1})d\bx_{k-1}.
\end{split}
\end{equation}
The update step involves obtaining the posterior density by conditioning on $Z_k$ and computing Bayes' rule as 
\begin{equation}\label{eq:bayes_rfs}
    p_{k|k}(\bx_k|Z_{1:k}) = \frac{\eta_k(Z_k|\bx_k)p_{k|k-1}(\bx_k|Z_{1:k-1})}{\int\eta_k(Z_k|\bx_k)p_{k|k-1}(\bx_k|Z_{1:k-1})d\bx_k}.
\end{equation}
%
%
%
%
The linear Gaussian assumption allows for the use of the closed-form solution to the RFS single-target Bayes' recursion in~\eqref{eq:chapmankolmogorov_rfs} and \eqref{eq:bayes_rfs} as proposed in~\cite[Proposition 4]{vo_bayesian_2008}.
Under the assumption of linear Gaussian process and measurement models, at most one target-generated measurement, state-independent probability of target detection, and uniform clutter, this closed-form solution reduces to the Gaussian mixture filter~\cite{5089549}. This type of filter is advantageous in target tracking frameworks since it enables managing multiple DA hypotheses by maintaining a set of possible tracks. Each hypothesis is represented by a Gaussian component in the posterior mixture. As new measurements are received, the algorithm updates the weights and parameters of these Gaussian components using Bayes' rule.
Under these conditions, if we let $m$ index the predictive hypotheses and $m'$ index the posterior hypotheses, the predicted density takes the form of a Gaussian mixture as
\begin{equation}\label{eq:predictive_distro}
    p_{k|k-1}(\bx_k|Z_{1:k-1}) = \sum_{m=1}^{M_{k|k-1}}\tilde{w}_{k|k-1}^{(m)}\mathcal{N}(\bx_k;\bbm_{k|k-1}^{(m)},\bP_{k|k-1}^{(m)}),
\end{equation}
where $M_{k|k-1}$ is the number of hypotheses at time $k$ for the prediction step, $\tilde{w}_{k|k-1}^{(m)}$ are the normalized predicted weights, and $\bbm_{k|k-1}^{(m)}$ and $\bP_{k|k-1}^{(m)}$ are the predicted mean and covariance of the $m$-th predicted hypothesis.
If this density is propagated through the likelihood model in~\eqref{eq:likelihood_rfs}, the resulting density is also a Gaussian mixture given by
\begin{equation}
    p_{k|k}(\bx_k|Z_{1:k}) = \sum_{m'=1}^{M_{k|k}} \tilde{w}_{k|k}^{(m')} \mathcal{N} ( \bx_k; \bbm_{k|k}^{(m')}, \bP_{k|k}^{(m')}),
\end{equation}
where $M_{k|k}$ is the number of hypotheses at time $k$ for the update step, $\tilde{w}_{k|k}^{(m')} = w_{k|k}^{(m')} / \sum_{m'=1}^{M_{k|k}} w_{k|k}^{(m')}$ are the normalized posterior weights, and $\bbm_{k|k}^{(m')}$ and $\bP_{k|k}^{(m')}$ are the posterior mean and covariance of the $m'$-th posterior hypothesis.
For the derivation and full expression of the predicted and updated means and covariances, see~\cite[Propositions 3 and 4]{vo_bayesian_2008}. Background on the standard results for the closed-form solution in the linear Gaussian case may be found in~\cite{Ristic2004BeyondTK,1105763}.

\begin{figure*}[h]
    
    \hspace{-0.4cm}\subfloat[Adaptation to time-varying jammer-induced bias.]{
        \includegraphics[width=1.05\textwidth]{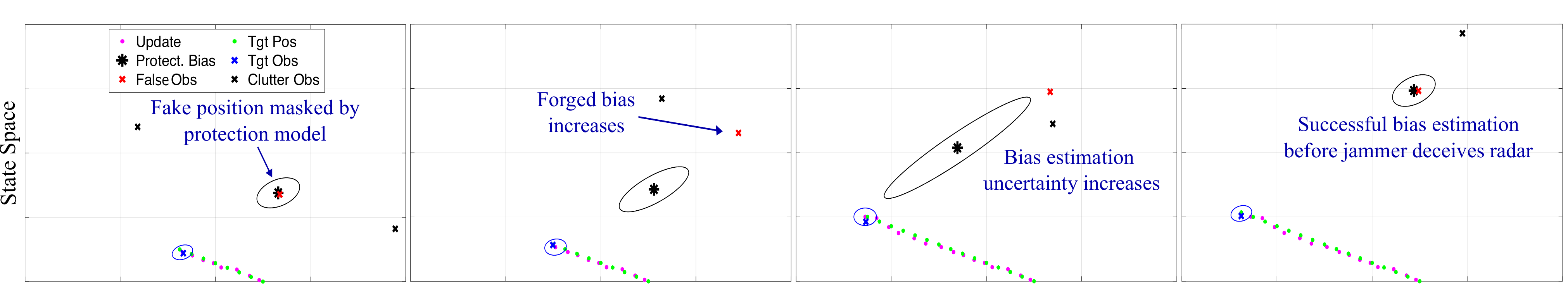}
        \label{fig:sub1}
    }
    \vspace{-0.01cm}

    \hspace{-0.4cm}\subfloat[Adaptation to time-varying attack status.]{
        \includegraphics[width=1.05\textwidth]{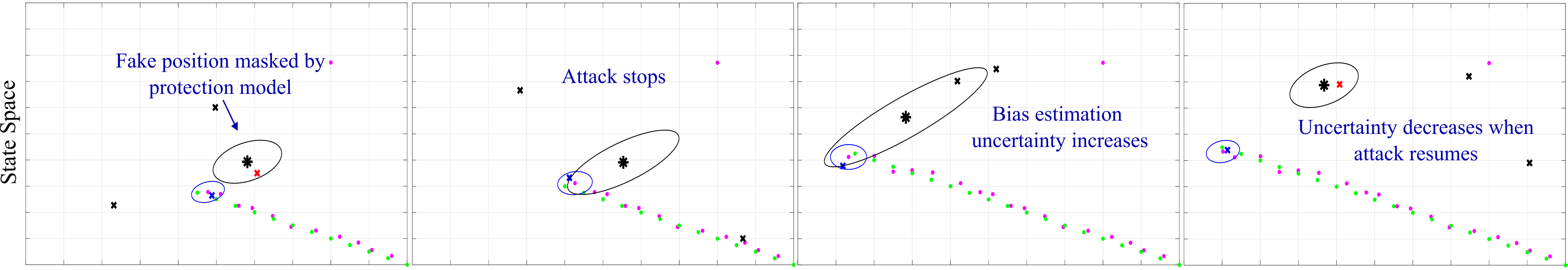}
        \label{fig:sub2}
    }
    \caption{Adaptive tracking in the presence of deception jamming. Each sequence consists of four frames showing the progression over time of the tracking process and the effectiveness of the protection model. In (a), the adaptive tracker compensates for a sudden increase in the jammer-induced bias. In (b), the adaptive tracker increases bias estimation uncertainty when the attack stops and reduces this uncertainty when the attack resumes.}
    \label{fig:progression_adaptive_tracking}
\end{figure*}
\subsection{Adaptive Estimation of Interference Bias}
The method presented in~\cite{vo_bayesian_2008} assumes a predetermined bias which, if incorrect, can lead to significant performance degradation comparable to that of the naive tracker shown in Figure~\ref{fig:intro}.
%
Figure~\ref{fig:progression_adaptive_tracking} illustrates the necessity of adaptive tracking in two distinct scenarios. 
\subsubsection{Strategy Against Single-Return Attacks}
To provide a robust and adaptive tracking solution, the proposed method dynamically estimates the interference bias $b_k$ by augmenting the state vector as $\tilde{\bx}_k = [\bp_k^\top, \bv_k^\top, b_k]^\top$.
The state vector is described by the linear Gaussian transition model in~\eqref{eq:transition_model} with parameters
\begin{equation}\hspace{-0.38cm} \label{eq:matrix_motion_model}
    \bF_k = \begin{bmatrix}
\bI_2 & \Delta\bI_2 & \mathbf{0}_{2\times 1} \\
\mathbf{0}_{2\times2} & \bI_2 & \mathbf{0}_{2\times 1} \\
\mathbf{0}_{1\times 2} & \mathbf{0}_{1\times 2}  & 1
\end{bmatrix}, 
\bQ_k = \sigma_q^2    \begin{bmatrix}
\frac{\Delta^4}{4} \bI_2 & \frac{\Delta^3}{2} \bI_2 & \mathbf{0}_{2\times 1}  \\
\frac{\Delta^3}{2} \bI_2 & \Delta^2 \bI_2 & \mathbf{0}_{2\times 1}  \\
\mathbf{0}_{1\times 2}   &\mathbf{0}_{1\times 2}   & \alpha \Delta
\end{bmatrix}, 
\end{equation}
where $\bI_m$ is the $m \times m$ identity matrix, $\mathbf{0}_{m \times n}$ is an $m \times n$ zero matrix, $\Delta$ is the radar sampling period, $\sigma_q$ is the process noise standard deviation, and $\alpha$ scales the uncertainty of the bias estimate.
The target position is observed through the model in~\eqref{eq:measurement_likelihood} with parameters
\begin{equation}\label{eq:measurement_model_params}
    \bH_k = \begin{bmatrix}\bI_2& \mathbf{0}_{2\times3}\end{bmatrix},\quad \bR_k = \sigma^2_r\bI_2,
\end{equation}
being $\sigma_r$ the measurement noise standard deviation.
The target trajectories described in Figure~\ref{fig:trajectories} include turns, for which a turning rate model is used for trajectory generation. However, the tracker consistently assumes the constant velocity model in~\eqref{eq:matrix_motion_model}. While this creates a model mismatch, this challenge is shared by all techniques tested in this study, including the benchmark.

The TTR models the jammer-generated measurements as in~\eqref{eq:likelihood_ck1}, with specific parameter choices designed to integrate the spatial characteristics of RGPO attacks. The observation matrix depends on the LOS vector estimate, which changes for each component of the Gaussian mixture in~\eqref{eq:predictive_distro}. For the $m$-th hypothesis, with predicted mean $\bbm_{k|k-1}^{(m)}$, the jammer observation model parameters are
\begin{equation} \label{eq:adaptive}
  \bB_k^{(m)} = \begin{bmatrix}
 \bI_2& \mathbf{0}_{2\times2}&\hat{\br}_k^{(m)}
\end{bmatrix}, \quad \bD_k = \bR_k,
\end{equation}
where $\hat{\br}^{(m)}_k = \frac{\bbm_{k|k-1}^{(m)} - \bp_k^0}{\norm{\bbm_{k|k-1}^{(m)} - \bp_k^0}}$.
%
%
Note that the bias vector is embedded in the model as $\bb_k^{(m)} = b_k\hat{\br}^{(m)}$ but does not appear as an additive term as shown in~\eqref{eq:likelihood_ck1}. This also applies to the extension to multi-pulse attacks.
%
%
%
\subsubsection{Extension to Multi-Pulse Attacks}
The proposed adaptive solution can seamlessly handle multiple attacks without additional algorithmic modifications when they start at similar times and exhibit similar pull-off velocities. This capability stems from the Poisson RFS clutter model assumed by the TTR, which inherently accounts for multiple detections. However, an extension of the model is required to maintain resilience when the parameters between RGPO attacks differ.
To address this challenge, the intensity function in~\eqref{eq:intensity_jammer} is modified to become a Gaussian mixture with a time-varying number of components $C_k$ as
\begin{equation} \label{eq:multi_intensity_function}
\lambda_{J,k}(\bz_k|\bx_k) = \sum_{i=1}^{C_k} \bar{\lambda}_{i,k} c_{i,k}(\bz_k|\bx_k)
\end{equation}
\begin{equation}
c_{i,k}(\bz_k|\bx_k) = \mathcal{N} \left( \bz_k; \bB_{i,k} \bx_k, \bD_k \right).
\end{equation}
Analogous to~\eqref{eq:adaptive}, $\bD_k = \bR_k$, and the jammer observation matrix for the $m$-th hypothesis can be expressed as
\begin{equation} \label{eq:adaptive_multiple}
  \bB_{i,k}^{(m)} = \begin{bmatrix}
 \bI_2 & \mathbf{0}_{2\times2} & \mathbf{0}_{2\times{(i-1)}} & \hat{\br}_k^{(m)} & \mathbf{0}_{2\times{(C_k-i)}}
\end{bmatrix}.
\end{equation}
Note that the model in~\eqref{eq:intensity_jammer} corresponds to the case when $C_k = 1$.
The complete clutter measurement model likelihood is given by
\begin{equation}\label{eq:multiple_model}
c_k(\bz_k|\bx_k) = w_{0,k} u(\bz_k) + \sum_{i=1}^{C_k} w_{i,k} c_{i,k}(\bz_k|\bx_k),
\end{equation}
where $w_{i,k} = \bar{\lambda}_{i,k}/\sum_{j=1}^{C_k} \bar{\lambda}_{j,k}$ for $i = 0, 1, \ldots, C_k$. Each mixture component is responsible for either mitigating an already detected attack or remaining vigilant for the potential appearance of a new attack.
%
%
%
Given that the number of attacks in the scene is unknown, we introduce a novel approach to dynamically manage $C_k$ to account for the unpredictable nature of RGPO attacks. This approach is used in Experiment 2.

We initialize the algorithm with one Gaussian component, i.e., $C_k=1$, in \textit{vigilant} status and $\tilde{\bx}_{1,k} = [\bp_k^\top, \bv_k^\top, b_{1,k}]^\top$ as the state vector. If the uncertainty in the estimation of bias $b_{1,k}$ remains below the threshold $U_\text{act}$ for $T_\text{act}$ time steps, it indicates that the tracker is confident about the estimated jammer-induced bias. Since uniform clutter would not consistently reduce uncertainty around a specific bias value, this suggests that the first component of the mixture is indeed tracking an RGPO attack. Considering this, the status of the first component is changed to \textit{active} and a new component in vigilant status is added to the mixture. This ensures that there is always a component prepared to handle potential new attacks. When a new $i$-th component is added, the state vector is augmented to estimate a new bias as $\tilde{\bx}_{i,k} = [(\tilde{\bx}_{(i-1),k})^\top, b_{i,k}]^\top$.
To save computational cost, DA hypotheses are reduced by sending a component to \textit{dormant} status if uncertainty in the estimation of its associated bias exceeds the threshold $U_\text{dorm}$ for $T_\text{dorm}$ time steps. The state vector is then reduced by removing the corresponding bias term and the mean and covariances of all posterior components are adjusted accordingly.
%
%
%
%
In Experiment 1, the number of Gaussian components is fixed at \(C_k = 1\) since this experiment focuses on scenarios 1 and 2, where only one attack can occur at a time. However, to account for the possibility of multiple non-simultaneous attacks, the Gaussian component is allowed to restart to the prior distribution if it appears to be in dormant status.

%
%
%
%
\begin{figure*}[h]
    \hspace{-1cm}
    \subfloat[Scenario 1: Straight line trajectory in the presence of a single RGPO attack at a time.]{
        \includegraphics[width=1.1\textwidth]{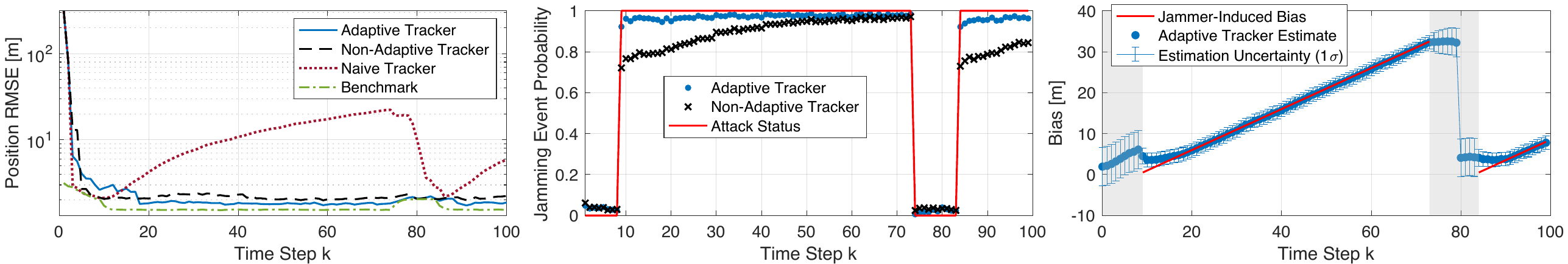}
        \label{fig:sub1}
    }
    \vspace{-0.1cm}
    
\hspace{-1cm}
    \subfloat[Scenario 2: Trajectory with three 3g turns in the presence of a single RGPO attack at a time.]{
        \includegraphics[width=1.1\textwidth]{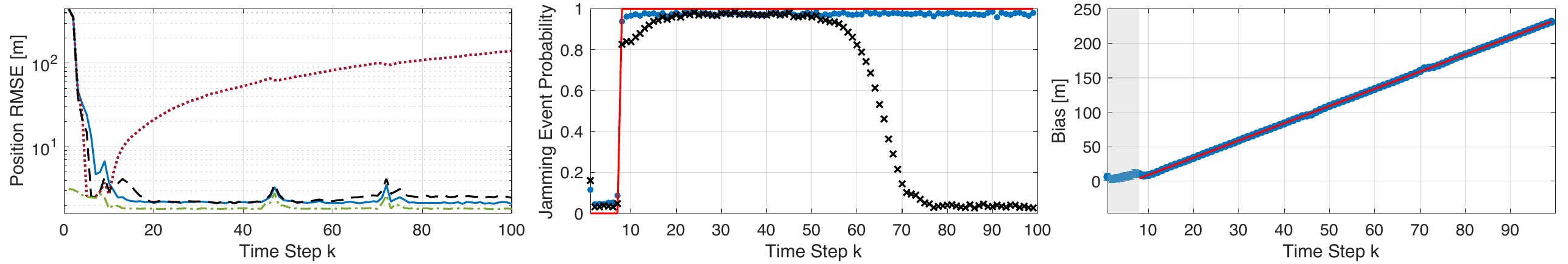}
        \label{fig:sub2}
    }
    \caption{Results obtained for Experiment 1 in the presence of a single RGPO attack in terms of position RMSE, probability of jamming event, and bias estimation. The grey area in the bias plot denotes when no attack is present.}
    \label{fig:results:exp1}
\end{figure*}
\vspace{-0.5cm}
\subsection{Non-Adaptive Highly Uncertain Approach}
We present an alternative that does not require state augmentation and provides a reasonable solution when accurate priors on the jammer-induced ranges are available.
The choice of parameters for the non-adaptive strategy is 
%
%
\vspace{-0.75cm}
\begin{equation}
   \bB_{\text{NA},k} = \begin{bmatrix}\bI_2& \mathbf{0}_{2\times 2}\end{bmatrix},\quad \bD_{\text{NA},k}^{(m)} = \hat{\bU}_k^{(m)}\boldsymbol{\Lambda} (\hat{\bU}_k^{\top})^{ (m)}. 
\end{equation}
In contrast to the adaptive strategy, which embeds the bias vector in the model by including the LOS vector in the observation matrix, the non-adaptive strategy adds the bias vector as a separate term as shown in~\eqref{eq:likelihood_ck1}. Specifically, $\bb_{\text{NA}, k}^{ (m)} = b_\text{NA}\hat{\br}_k^{ (m)}$, where $b_\text{NA}$ is the predetermined jammer-induced bias assumed by the tracker.
Due to this modeling difference, the covariance matrix must be flattened along the LOS direction by incorporating $\hat{\br}_k^{ (m)}$ into the matrix of eigenvectors \(\hat{\mathbf{U}}_k^{ (m)}\), with $\boldsymbol{\Lambda}$ being an unbalanced diagonal matrix that introduces higher uncertainty along the LOS.
%


RGPO attacks typically start with negligible bias and then gradually increase the induced ranges to divert attention away from the TOI. Considering this, $b_\text{NA}$ should be set so that, when added to the $3\sigma$ bound along the LOS direction provided by $\mathbf{D}_{\text{NA},k}^{ (m)}$, it covers the vicinity of the TOI.
In this way, the protection model is prepared to mitigate the RGPO attack as soon as it starts.
%


\vspace{-0.4cm}

\subsection{Jamming Detection}
To detect the presence of the deception jammer, we calculate the probability of the event \( A = \{|J_k(\bx_k)| \geq 1\} \), which represents the scenario where at least one of the measurements in \( Z_k \) is identified by the tracker as being jammer-generated. If this condition is met, we infer that the jammer is present.
%
%
A probability value is calculated at the update step for each posterior hypothesis.
%
The set of measurements assumed as secondary for the posterior hypothesis $m'$ is denoted as $S_k^{(m')} = \{\bs_{k,1}, \hdots, \bs_{k,|S_k^{(m')}|}\}$, and corresponds to $S_k^{(m')} = Z_k$ under target misdetection hypothesis and to $S_k^{(m')} = Z_k\backslash\{\bz_k^{(m')}\}$ under the hypothesis that the target is detected with measurement $\bz_k^{(m')}$.
To calculate \( P(A^{(m')} \coloneqq \{|J_k^{(m')}(\bx_k)| \geq 1\}) \), we first determine the probability that all secondary measurements are considered false alarms by the TTR, and then take its complement as
\begin{equation}
\begin{split}
    P(A^{(m')}) & = 1 - P(S_k^{(m')}\in W_k) 
    = 1- \prod_{j=1}^{|S_k^{(m')}|} P( \bs_{k,j}^{(m')} \in W_k ).
    \end{split}
\end{equation}
%
%
Here, we use the fact that false alarm detections are independent, given that \( W_k \) is a Poisson RFS with the intensity function in~\eqref{eq:likelihood_clutter}.
Each factor in the product can be computed by Bayes' rule as
\begin{equation}
\begin{split}
     P( \bs_{k,j}^{(m')} \in W_k ) &= \frac{w_{0,k}u(\bs_{k,j}^{(m')})}{\int c_k(\bs_{k,j}^{(m')}|\bx_k)\mathcal{N}(\bx_k;\bbm_{k|k-1}^{(m)},\bP_{k|k-1}^{(m)})d\bx_k},
    \end{split}
\end{equation}
%
%
%
where the denominator represents the total clutter probability for the measurement \(\bs^{(m')}_{k,j}\) under the posterior hypothesis \(m'\). Since this hypothesis is formed by updating the predicted hypothesis \(m\), the denominator includes the density of the \(m\)-th component of the predicted distribution shown in~\eqref{eq:predictive_distro}. This integral can be computed as
\begin{equation}
\begin{split}
   &\int c_k(\bs_{k,j}^{(m')}|\bx_k)p(\bx_k)d\bx_k 
    = w_{0,k}u(\bs_{k,j}^{(m')}) \\&+ \sum_{i=1}^{C_k}w_{i,k}\int c_{i,k}(\bs_{k,j}^{(m')}|\bx_k)p(\bx_k)d\bx_k \\&= w_{0,k}u(\bs_{k,j}^{(m')}) + \sum_{i=1}^{C_k}w_{i,k}\mathcal{N}(\bs_{k,j}^{(m')};\bm{\mu}_{i,k}^{(m)}, \bSigma_{i,k}^{(m)}),
\end{split}
\end{equation}
where the density $p(\bx_k) = \mathcal{N}(\bx_k;\bbm_{k|k-1}^{(m)},\bP_{k|k-1}^{(m)})$ is used as a notation shorthand, $\bm{\mu}_{i,k}^{(m)} = \bB_{i,k}^{(m)}\bbm_{k|k-1}^{(m)}$, and $\bSigma_{i,k}^{(m)} = \bB_{i,k}^{(m)}\bP_{k|k-1}^{(m)}(\bB_{i,k}^{(m)})^\top + \bD_k$.
This expression has been derived using the model from the adaptive strategy for multi-pulse attacks. An analogous expression without the summation over the \(C_k\) mixture components is obtained for the non-adaptive implementation.
\subsection{Implementation Notes}
%
%
%
If the posterior at time \(k-1\) has \(M_{k-1}\) mixture components, then the posterior at time \(k\) has $M_k$ components as~\cite{vo_bayesian_2008}
\begin{equation}
     M_{k-1} \left( C_k|Z_k| + |Z_k|C_k^{|Z_k|-1} \right) = \mathcal{O} \left( M_{k-1} \cdot C_k^{|Z_k|} \right),
\end{equation}
where we have included $C_k$, the varying number of Gaussian components from the extension to multiple RGPO attacks in~\eqref{eq:multi_intensity_function}.
The closed-form solution does not guarantee tractability, given that the increase of mixture components is unbounded~\cite{CONG19991}. We resort to pruning and capping as approximation techniques to manage the number of components, although more sophisticated strategies exist  \cite{Murphy2001}.
Pruning removes mixture components with low weights while capping limits the total number of components by keeping only the ones with highest weights.
%
%
%
%
%
To prevent the mirror effect, where the TTR mistakes the TOI for the jammer, we can leverage domain knowledge about RGPO attacks.
By using spatial gating, we set the weights of hypotheses with a negative bias to zero.
\vspace{-0.4cm}
\section{SIMULATION RESULTS}
%

%
This section outlines the experimental setup and discusses results in terms of position RMSE, jamming detection accuracy, and the ability to handle multiple RGPO attacks. A LoE study is included to verify that the proposed strategy works effectively in the absence of interference.
\vspace{-0.45cm}
\subsection{Experimental Setup}

Four simulations are conducted with the scenarios presented in Section II.B (see Figure~\ref{fig:trajectories}), spanning 100 time steps with $\Delta = 0.5$ s.
The expected number of false alarm detections is given by $\bar{\lambda}_{0} = \lambda_{0}V$, where $\lambda_{0} = 2 \times 10^{-5}$ m$^{-2}$ and $V$ is the volume of $\mathcal{Z}$.
%
The observation region, in units of meters, is defined as $\mathcal{Z} = [0,1000] \times [0,1000]$, with a balanced probability of detection between target and jamming returns as $p_D = p_J = 0.98$.
For all scenarios, the observation model is the one in~\eqref{eq:measurement_likelihood}, with the parameters defined in~\eqref{eq:measurement_model_params} and $\sigma_r = \sqrt{5}$ m. The TTR assumes the constant velocity model in~\eqref{eq:transition_model} with the parameters in~\eqref{eq:matrix_motion_model} and $\sigma_q = \sqrt{5}$ m in Scenarios 1 and 3, and $\sigma_q = \sqrt{40}$ m in Scenarios 2 and 4. The latter accounts for higher accelerations during turns, where the constant velocity assumption poses a challenge. Nevertheless, the process noise used for trajectory generation is set to zero, except during time steps when the target performs a 3 g turn. This introduces a model mismatch that is further discussed when we present the benchmarks at the end of this subsection.
For attack generation, Experiment 1 uses pull-off velocities as defined in \eqref{eq:rgpo} of \(0.5\ \text{m/s}\) for the two attacks in Scenario 1 and \(5\ \text{m/s}\) for the single attack in Scenario 2. In Experiment 2, for Scenarios 3 and 4, the velocities are \(5\ \text{m/s}\) for the first attack and \(3\ \text{m/s}\) for subsequent attacks.
We conduct 1000 Monte Carlo runs on the same target trajectory, with independently generated measurements for each realization.
We perform pruning at each time step with a weight threshold of $10^{-5}$ and capping by limiting the number of hypotheses to $100$.

For the adaptive strategy, the uncertainty of the bias estimate is scaled by a factor of $\alpha=10$ as specified in~\eqref{eq:matrix_motion_model}.
Initialization of the filter is done with the prior $p_0 = \mathcal{N}(\cdot; [500, 500, 0, 0, 0], \text{diag}(1e4, 1e4, 1e2, 1e2, 500))$. In Experiment 1, the assumed protection parameter in \eqref{eq:intensity_jammer} is set to $\bar{\lambda}_{1} = 3$, and the tracker uses the model described in Section III.B.1.
In Experiment 2, the tracker uses the model described in Section III.B.2, which includes the extension against multi-pulse attacks, and the assumed protection parameter remains $\bar{\lambda}_{i} = 3$ for each \(i\)-th component. When a new component is created, its prior mean and variance are the ones for the bias in $p_0$, i.e., with a mean $0$ and a variance of $500$. The activation thresholds are set to \(U_\text{act} = 5\ \text{m}\) in bias uncertainty and \(T_\text{act} = 7\ \text{s}\), while the deactivation thresholds are set to \(U_\text{dorm} = 5\ \text{m}\) and \(T_\text{dorm} = 4\ \text{s}\). The same deactivation thresholds $U_\text{dorm}$ and $T_\text{dorm}$ are used in Experiment 1 to restart the single Gaussian component when appearing dormant. 
\begin{figure*}[h]
    \hspace{-1cm}
    \subfloat[Scenario 3: Straight line trajectory in the presence of multiple, simultaneous RGPO attacks.]{
        \includegraphics[width=1.1\textwidth]{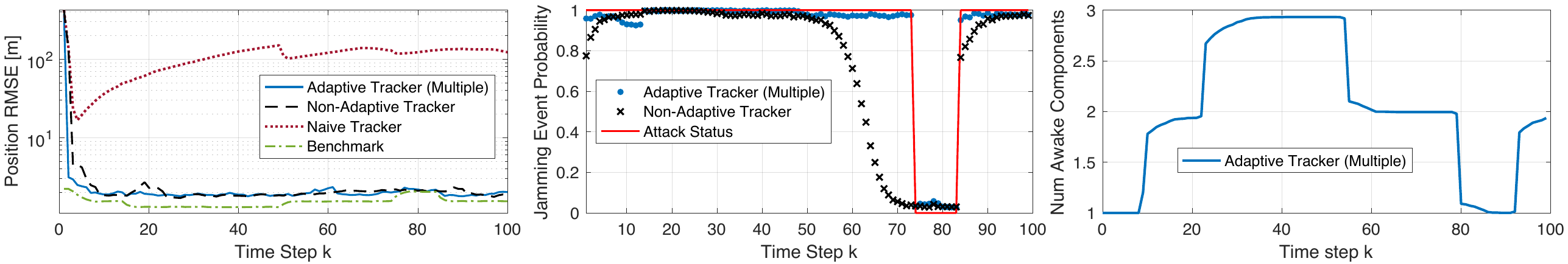}
        \label{fig:sub1}
    }
    \vspace{-0.1cm}
    
\hspace{-1cm}
    \subfloat[Scenario 4: Trajectory with three 3g turns in the presence of multiple, simultaneous RGPO attacks.]{
        \includegraphics[width=1.1\textwidth]{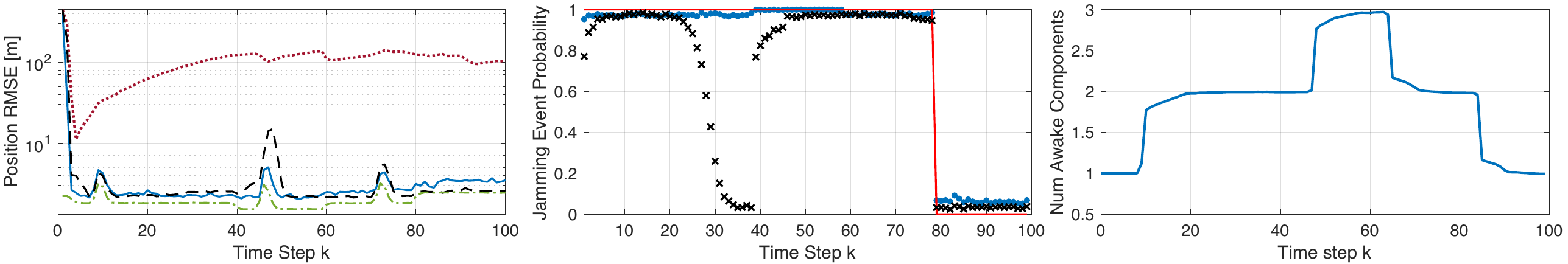}
        \label{fig:sub2}
    }
    \caption{Results obtained for Experiment 2 in the presence of multiple RGPO attacks in terms of position RMSE, probability of jamming event, and number of awake Gaussian components $C_k$ following the model in~\eqref{eq:multiple_model}.}
    \label{fig:results:exp2}
\end{figure*}

For the non-adaptive strategy, the tracker uses the model described in Section III.C with \(b_\text{NA} = 70\ \text{m}\), \(\boldsymbol{\Lambda} = \text{diag}(500,1)\) and prior density $p_{\text{NA},0} = \mathcal{N}(\cdot; [500, 500, 0, 0], \text{diag}(1e4, 1e4, 1e2, 1e2))$.
The eigenvalue corresponding to the LOS direction matches the initial variance for bias estimation in the adaptive approach, i.e., 500.
The adaptive method starts with a variance of 500 and adjusts it based on the occurrence of attacks, while the non-adaptive method maintains a high level of uncertainty under all conditions since it lacks adaptive capabilities to estimate the bias.
%

%

%
%
%

%
As a benchmark, we use a clairvoyant tracker without DA uncertainty operating under a constant velocity model. Although prone to errors during high-acceleration turns, this benchmark establishes a performance bound for the proposed technique by demonstrating its potential when RGPO attacks and false alarms are successfully identified. When attack identification is successful, jamming detections become additional observations that enhance tracking performance.
%
Due to the model mismatch between the process noise assumed by the TTR and the actual noise in the generated trajectories, the PCRB is not used as a benchmark in the first two experiments. Instead, we use the clairvoyant tracker, which aligns with our focus on evaluating TTR performance under the constant velocity assumption. In the LoE experiments, where we set $\sigma_q = 0$ both at the TTR and for trajectory generation, the PCRB is used as a benchmark.
In all the experiments, we also include a naive tracker that only accounts for false alarms as \( K_k(x_k) = W_k \). This unprotected model demonstrates the performance of a single-target tracker that handles DA uncertainties for uniform clutter but is oblivious to RGPO attacks, hence the term \textit{naive}.
\vspace{-0.4cm}

\subsection{Experiment 1: Single RGPO Attack}
Results for Experiment 1 are presented in Figure~\ref{fig:results:exp1}.
In terms of position RMSE, both the adaptive and non-adaptive approaches demonstrate near-optimal performance, as indicated by the benchmark in both scenarios. The adaptive tracker achieves slightly lower errors than the non-adaptive tracker after an initial adaptation period of $20$ time steps. In Scenario 2, both strategies recover within approximately five time steps following the disruptions introduced by the 3 g turns. The deception jammer walks the naive tracker away, while the proposed methods maintain a lock on the TOI, improving accuracy by up to $20$ meters.

When it comes to jamming detection performance, the advantage of the adaptive approach over its non-adaptive counterpart is particularly evident. The non-adaptive strategy struggles with biases that exceed the $3\sigma$ bound of the protection model covariance, making the TTR unaware of the attack. This is especially noticeable for \(k > 60\) in Scenario 2, where jammer-induced biases exceed $100$ meters and the calculated probability of the jammer rapidly drops to zero despite the ongoing attack. Nonetheless, this does not impact position RMSE, as the induced bias, although leaving the tracker unprotected, remains too distant from the TOI to disrupt the TTR lock.

The uncertainty in bias estimation for the adaptive approach increases in the absence of interference. This can be seen particularly in Scenario 1, given the scale of the figure. The increase in uncertainty occurs because no jammer observations are available, and therefore, no updates using the jammer return likelihood in~\eqref{eq:likelihood_ck1} are performed.

%


\vspace{-0.5cm}
\subsection{Experiment 2: Multiple RGPO Attacks}
Results for Experiment 2 are presented in Figure~\ref{fig:results:exp2}, where the subfigures on the right show the average number of \textit{awake} Gaussians, i.e., those in either vigilant or active status, for the adaptive strategy. In terms of position RMSE, both the adaptive and non-adaptive methods maintain errors close to the benchmark. Similar to Experiment 1, both algorithms recover after approximately five time steps following a maneuver, with the adaptive method providing more resiliency in the second turn. Since Scenarios 3 and 4 begin under the presence of a jammer, the adaptation period seen in Scenarios 1 and 2, where the adaptive strategy initially had slightly higher errors, is not observed. The multiple jammer observations help refine state estimates when the attack model is effectively integrated into the RFS clutter model. This is particularly evident in Scenario 4, where the benchmark achieves the lowest RMSE for $40 < k < 60$, the period during which two simultaneous attacks occur.
%
%
In terms of deception detection, the non-adaptive approach finds difficulty in detecting interference when jammer-induced biases deviate significantly from the assumed priors on jamming ranges, similar to what was observed in Experiment 1. For example, during time steps \(65 < k < 75\) in Scenario 3, the non-adaptive approach outputs a deception probability below 20\% despite the presence of interference. A similar issue arises in Scenario 4 for \(30 < k < 40\). In contrast, the adaptive strategy reliably detects deception.

\begin{figure}[h]
        \hspace{-0.35cm}
        \includegraphics[width=0.5\textwidth]{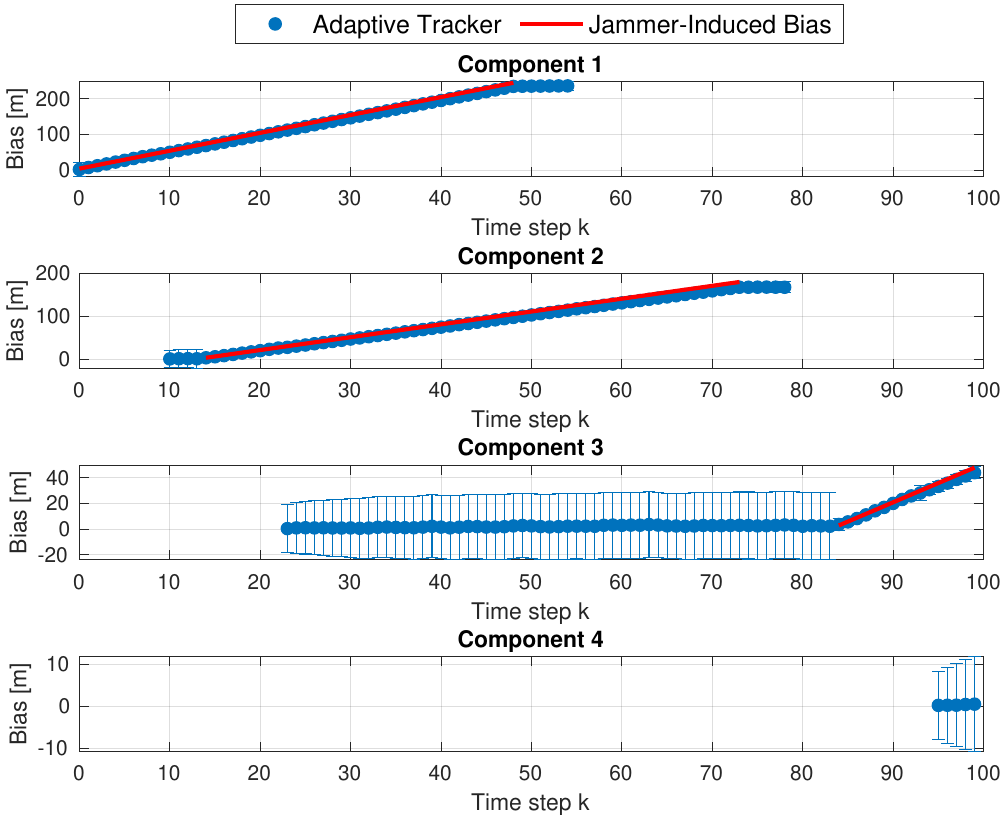}       
        \caption{Bias estimation results for the four components in the adaptive approach for multiple RGPO mitigation in Scenario 3.}\label{fig:bias_estimation_multiple}
\end{figure}
\vspace{-0.5cm}

Figure~\ref{fig:bias_estimation_multiple} analyzes bias estimation uncertainty for the adaptive strategy, helping to illustrate the dynamic management of $C_k$. Although one jammer is present from the start of the simulation, the initial Gaussian does not transition to active status until \(k > T_\text{act} = 7\). At that point, a second component is created in vigilant status, preparing for potential new attacks. In Scenario 3, the uncertainty of this second component decreases when a second attack appears at \(k = 15\). Once this uncertainty remains below \(U_\text{act}\) for more than \(T_\text{act}\) time steps, the second component transitions to active status, and a third component is created in vigilant status. When the first attack ends at \(k = 50\), $C_k$ decreases to $2$. Similar events occur throughout the remainder of the simulation and in Scenario 4, illustrating the dynamic management of \(C_k\), which adapts to the occurrence of attacks.


\vspace{-0.5cm}
\subsection{Loss of Efficiency}
\begin{figure}[h]
        \hspace{-0.35cm}
        \includegraphics[width=0.51\textwidth]{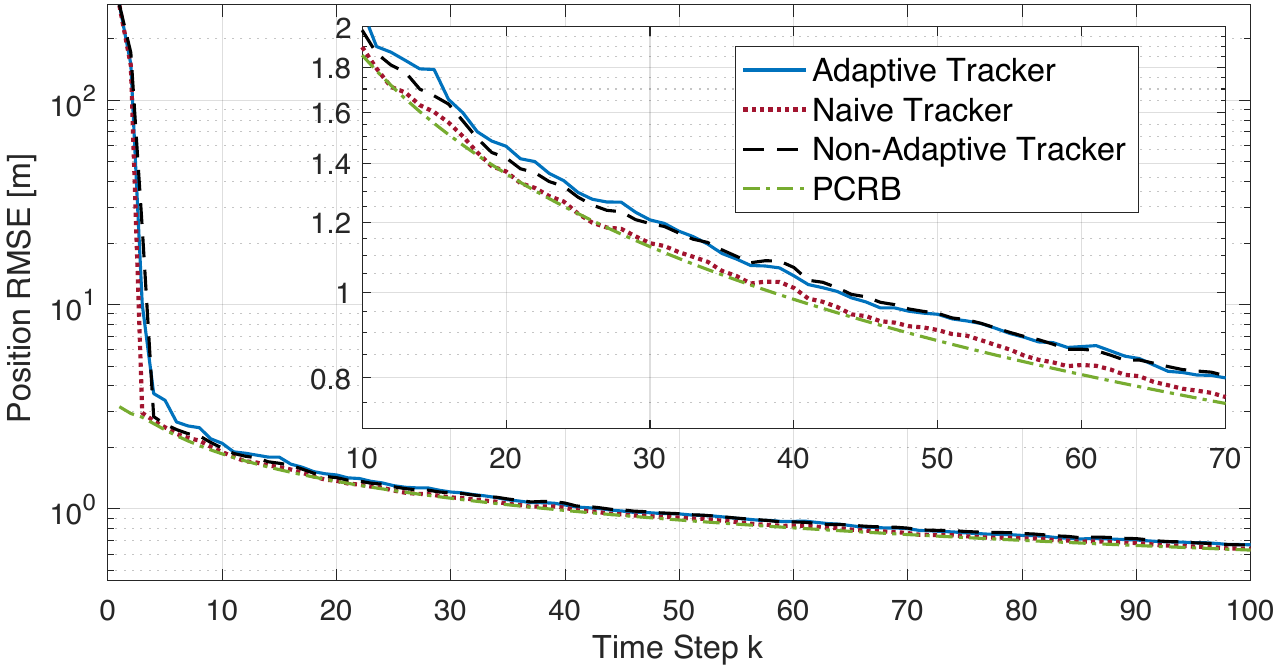}
        \caption{LoE analysis of the proposed methods in terms of position RMSE under nominal conditions (no attack).}\label{fig:loe_anlysis}
\end{figure}

In Figure~\ref{fig:loe_anlysis}, we present a LoE analysis in terms of position RMSE, with results compared to the lower bound as given by the PCRB. The naive tracker converges to the optimal bound since this experiment assumes nominal conditions (absence of attack). The proposed strategies demonstrate successful tracking, with the difference in position RMSE relative to the PCRB remaining almost negligible after \(k=10\).
%
%
%

%
%
\vspace{-1.43cm}
\section{CONCLUSION}

In this paper, we introduce a resilient radar target tracking framework that effectively counters main-beam range deception jamming attacks. We use random finite sets to model measurement sets with variable cardinality and apply multiple hypothesis tracking to address data association uncertainty. Our approach leverages motion state information and remains feature-independent by incorporating knowledge about the spatial behavior of range gate pull-off attacks into the clutter model.
We develop an adaptive solution that dynamically estimates jammer-induced biases, alongside a non-adaptive approach that performs well when accurate priors on deception ranges are available. Additionally, we introduce a novel method for jamming detection and a solution for managing the mixture components involved in a countermeasure against multi-pulse attacks.
In terms of position error, both the adaptive and non-adaptive strategies demonstrate near-optimal performance, improving accuracy by up to 20 meters when compared to an unprotected tracker. The adaptive approach shows a clear advantage in jamming detection due to its ability to reduce uncertainty in the interference spatial model. 
The proposed strategies maintain tracking performance in the absence of jamming, as shown in the loss of efficiency analysis using the posterior Cramér-Rao bound as a benchmark.
Overall, the introduced solutions maintain robust tracking in challenging environments, such as when the target is maneuvering with high accelerations while facing simultaneous RGPO attacks.






\section*{ACKNOWLEDGMENTS}

Research was sponsored by the Army Research Laboratory and was
accomplished under Cooperative Agreement Number W911NF-23-2-0014. The
views and conclusions contained in this document are those of the
authors and should not be interpreted as representing the official
policies, either expressed or implied, of the Army Research Laboratory
or the U.S. Government. The U.S. Government is authorized to reproduce
and distribute reprints for Government purposes notwithstanding any
copyright notation herein. This work has been partially supported by the National Science Foundation under Awards ECCS-1845833 and CCF-2326559.


\bibliographystyle{IEEEbib}%
\bibliography{ofdmradar_adversarial, mtt, dj_mht}
\vspace{-0.5cm}
\begin{IEEEbiography}[{\includegraphics[width=1in,clip,keepaspectratio]{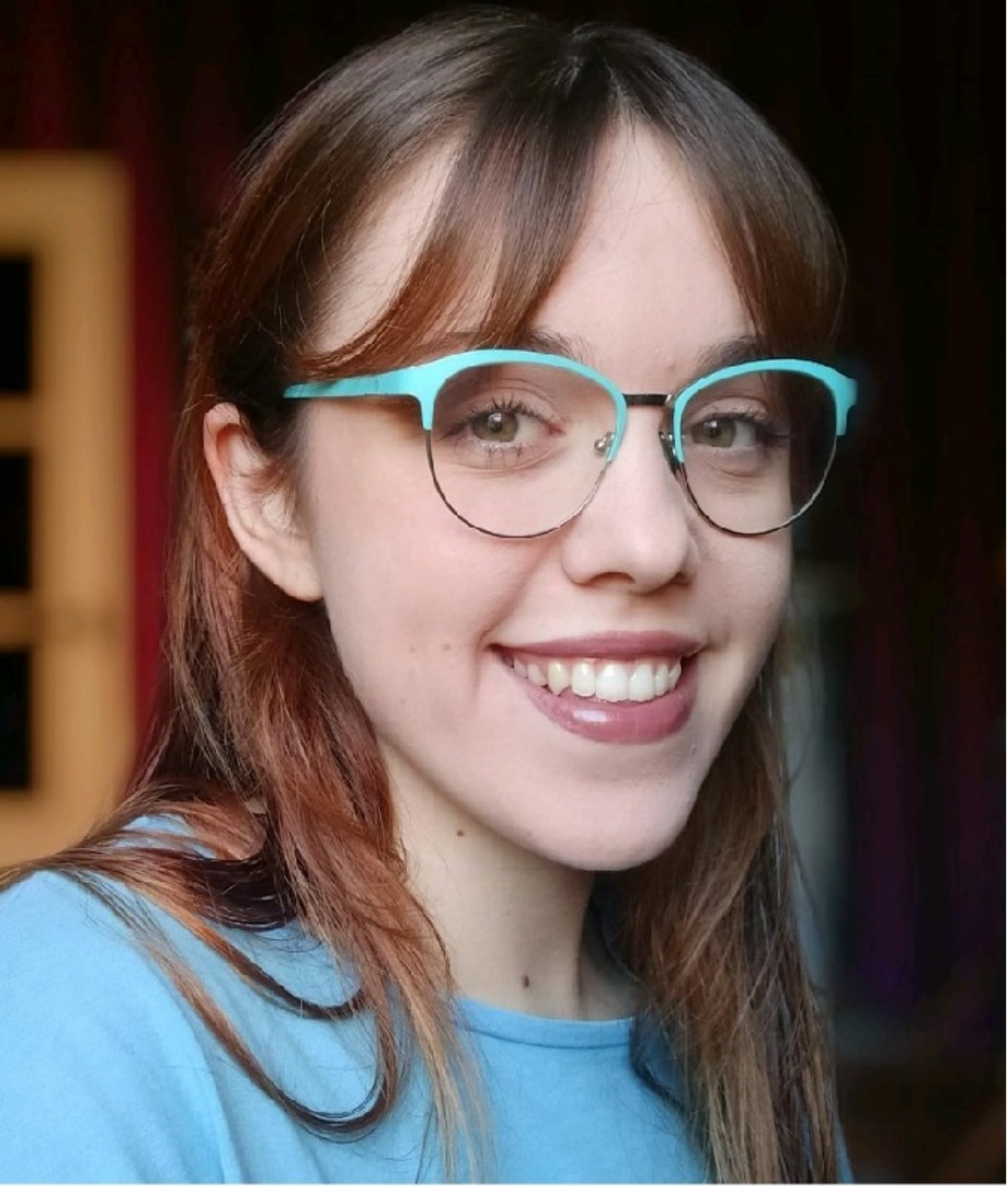}}]{Helena Calatrava}{\space}received her BS and MS degrees in Electrical Engineering from Universitat Politècnica de Catalunya, Barcelona, Spain, in 2020 and 2022, respectively. She is currently a PhD candidate in Electrical Engineering at Northeastern University, Boston, MA. Her research focuses on Bayesian filtering, physics-informed machine learning, anti-jamming and signal processing for GNSS and radar applications.
\end{IEEEbiography}%

\begin{IEEEbiography}[{\includegraphics[width=1in,clip,keepaspectratio]{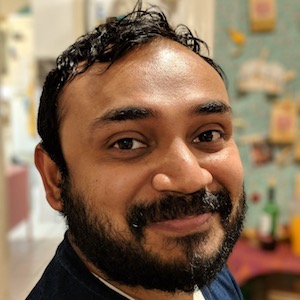}}]{Aanjhan Ranganathan}{\space} is
Associate Professor at
Northeastern University,
Boston. His research
revolves around the
security and privacy of
wireless networks with a strong focus on
autonomous cyber-physical systems and
smart ecosystems. He is a recipient of
several awards, including the prestigious
NSF CAREER award, the outstanding
dissertation award from ETH Zurich, the
regional winner of European Space
Agency’s Satellite Navigation competition,
and the Cyber Award from armasuisse
(Switzerland’s Department of Defense).
\end{IEEEbiography}%



\begin{IEEEbiography}
[{\includegraphics[width=1in,height=1.25in,clip,keepaspectratio]{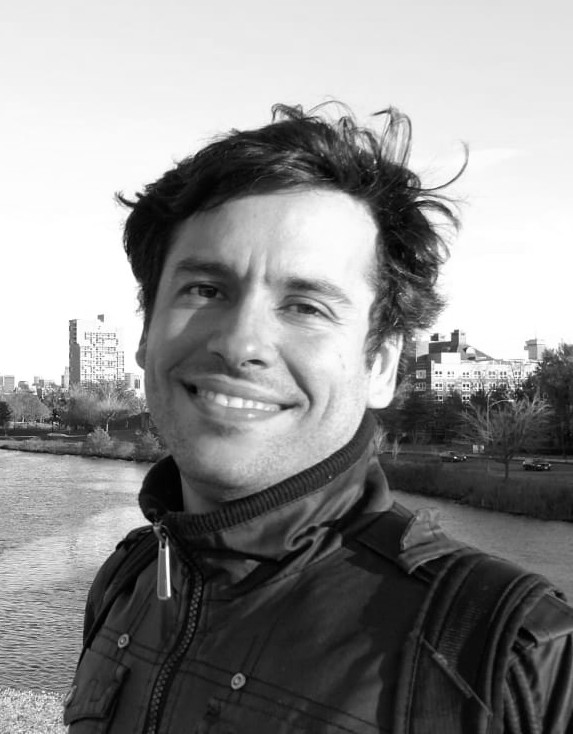}}]
{Tales Imbiriba} (Member, IEEE)   
is an Assistant Research Professor at the ECE dept., and Senior Research Scientist at the Institute for Experiential AI, both at Northeastern University (NU), Boston, MA, USA. He
received his Doctorate degree from the Department of Electrical Engineering (DEE) of the Federal University of Santa Catarina (UFSC), Florian\'opolis, Brazil, in 2016. He served as a Postdoctoral Researcher at the DEE--UFSC (2017--2019) and at the ECE dept. of the NU (2019--2021). 
His research interests include audio and image processing, pattern recognition, Bayesian inference, online learning, and physics-guided machine learning.
\end{IEEEbiography}

\begin{IEEEbiography}
[{\includegraphics[width=1in,height=1.25in,clip,keepaspectratio]{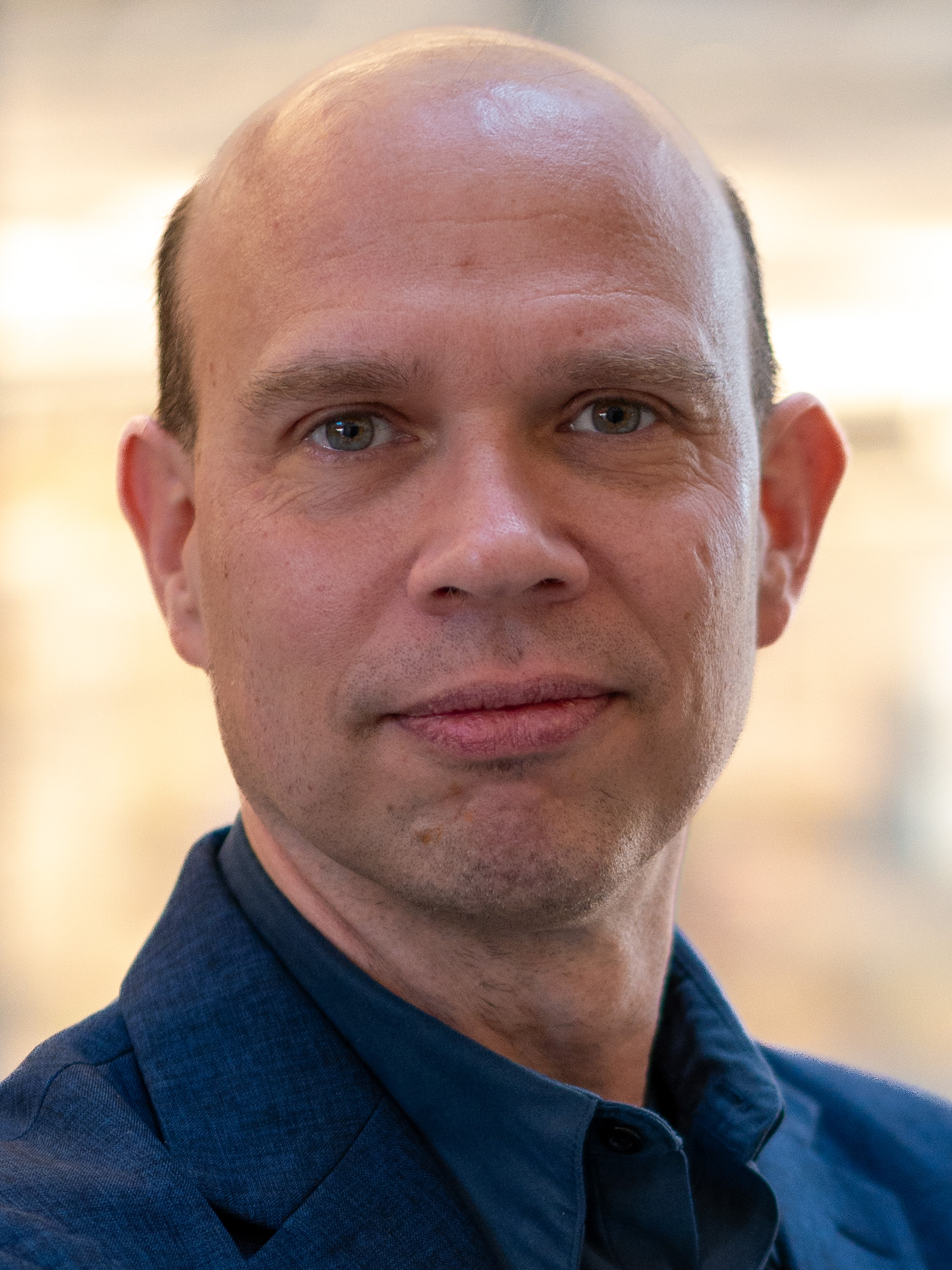}}]
{Gunar Schirner} (S'04--M'08) holds PhD (2008) and MS (2005) degrees in
  electrical and computer engineering from the University of California,
  Irvine. He is currently an Associate Professor in Electrical and
  Computer Engineering at Northeastern University. His research interests
  include the modelling and design automation principles for domain
  platforms, real-time cyber-physical systems and the
  algorithm/architecture co-design of high-performance efficient edge
  compute systems.
\end{IEEEbiography}
\vspace{-1em}

\begin{IEEEbiography}[{\includegraphics[width=1in,height=1.25in,clip,keepaspectratio]{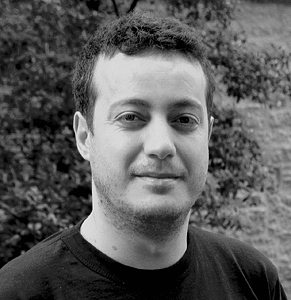}}]%
{Murat Akcakaya} (Senior Member, IEEE)
received his Ph.D. degree in Electrical Engineering from the Washington University in St. Louis, MO, USA, in December 2010. He is currently an Associate Professor in the Electrical and Computer Engineering Department of the University of Pittsburgh. His research interests are in the areas of statistical signal processing and machine learning.
\end{IEEEbiography}
\vspace{-1em}

\begin{IEEEbiography}[{\includegraphics[width=1in,height=1.25in,clip,keepaspectratio]{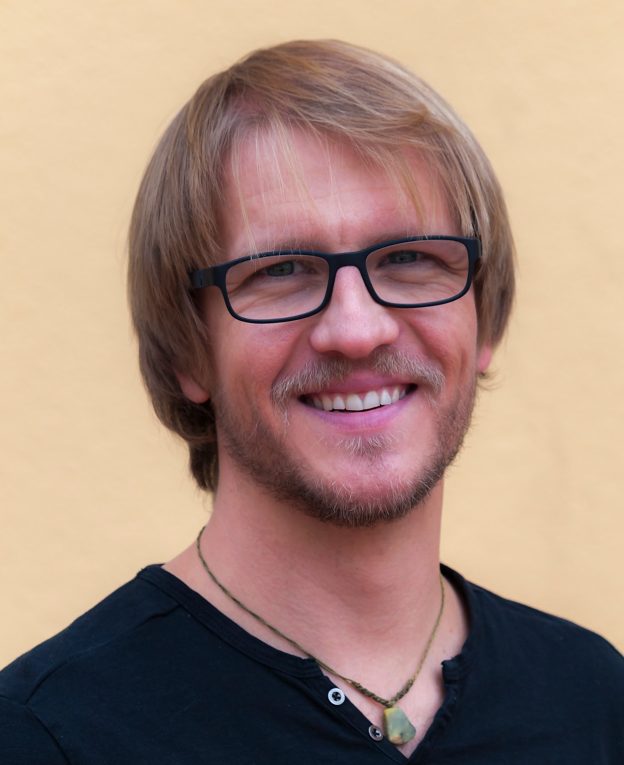}}]{Pau Closas}(Senior Member, IEEE),
is an Associate Professor in Electrical and Computer Engineering at Northeastern University, Boston MA.
He received the MS and PhD in Electrical Engineering from UPC in 2003 and 2009, respectively, and a MS in Advanced Mathematics from UPC in 2014. 
His primary areas of interest include statistical signal processing and machine learning, with applications to positioning and localization systems. 
\end{IEEEbiography}

\end{document}